\documentclass[beig,11pt]{article}
\usepackage{color}
\usepackage{amscd}
\usepackage[psamsfonts]{amssymb}
\usepackage{beig,transfig,amssymb,amsmath}
\newtheorem{thm}{Theorem}[section]
\newtheorem{propos}{Proposition}[section]
\newtheorem{lem}{Lemma}[section]

\newcounter{remarques}[section]
\renewcommand{\theremarques}{\arabic{section}.\arabic{remarques}}
\def\Rm{\refstepcounter{remarques}{\bf Remark \theremarques} }
\def\ali{\hfill\break}
\def\w{\omega}
\def\W{\Omega}

\def\demi{{1\over 2}}
\def\ddd{{\cal D}}

\def\nn{{<n>}}

\def\infi{{<\infty>}}

\def\ccc{{\cal C}}
\def\j1p0{{j_1^0,\ldots ,j_p^0}}

\def\jp1{{j_p,\ldots ,j_1}}
\def\j1p{{j_1,\ldots ,j_p}}

\def\BZ{{{\Bbb Z}}}
\def\BP{{{\Bbb P}}}
\def\Pp{{{\Bbb P}^2}}
\def\BN{{{\Bbb N}}}
\def\BR{{{\Bbb R}}}
\def\BC{{{\Bbb C}}}

\def\F1{{{F^{(1)}}}}

\def\supp{{{\hbox{supp}}}}

\def\Sl{{{\hbox{Sl}}}}
\def\im{{{\hbox{Im}}}}
\def\fff{{{\cal F}}}
\def\jjj{{{\cal J}}}
\def\nun{{{<n+1>}}}
\def\tr{{\hbox{Tr}}}

\begin{document}
\centerline{\Large \bf Spectral analysis of a self-similar}
\centerline{\Large\bf Sturm-Liouville operator}
%\ali\ali
% \centerline{\Large\bf Propri\'et\'es spectrales des r\'eseaux
% auto-similaires}
% \centerline{\Large\bf et it\'eration d'applications rationelles}
\ali \ali \centerline{\Large\bf Christophe Sabot} \ali
\begin{center}
Laboratoire de Probabilit\'es et mod\`eles al\'eatoires,
Universit\'e Paris 6,
\\
4, Place Jussieu, 75252 Paris cedex 5, \\
and
\\
Ecole Normale Sup\'erieure,
\\
DMA, 45, rue d'Ulm, 75005 Paris, \footnote{E-mail address:
sabot@ccr.jussieu.fr}
\end{center}
\ali \ali \ali {\bf Abstract:} In this text we  describe the spectral
nature (pure point or continuous) of a self-similar Sturm-Liouville
operator  on the line or the half-line.
 This is motivated by the more general problem of understanding
the spectrum of Laplace operators on unbounded finitely ramified
self-similar sets. In this context, this furnishes the first
example of a description of the spectral nature of the operator
in the case where the so-called "Neumann-Dirichlet" eigenfunctions
are absent.
  \ali \ali \ali {\bf AMS classification:}
34L10(34L20,82B44)
 \ali\ali\ali {\bf Key words:} Spectral theory,
Eingenfunctions expansion, Sturm-Liouville operators, dynamics in
several complex variables, analysis on
self-similar sets, fractals.

\vfill\break

In this text we consider a family of self-similar Sturm-Liouville
operators, indexed by a parameter $\w$ called the blow-up, and
describe the nature of the spectrum (pure point or continuous).
This study is motivated by the problem of understanding the nature
of the spectrum for the class of self-similar Laplacians on
finitely ramified self-similar sets. In \cite{Sabot3},
\cite{Sabot5}, \cite{Sabot-review1},
 we proved that the spectral properties of these
operators are related to the dynamics of a certain renormalization
map, which is a rational map of a smooth complex projective
variety. The main question emerging from these works is the nature
of the spectrum of these operators in the non-degenerate case
$d_\infty=N$ (cf \cite{Sabot5}, $d_\infty$ is the asymptotic
degree of the renormalization map, $N$ is the number of subcells
of the set at level 1). As far as the author knows, there
is no non-trivial example where the nature of the spectrum is
understood in this case.

The family of self-similar Sturm-Liouville operators we consider
here corresponds to the case $d_\infty=N$, and
we are able to determine, for all values of the blow-up $\w$,
whether the spectrum is pure point or continuous (we are not able
to distinguish the singular continuous and the absolute continuous
part). This example also illustrates \cite{Sabot6}, since it shows
that for atypical value of the blow-up $\w$ the spectral nature of
the operator can be radically different.

The proofs of the results suggest that the spectral nature of the
operators may be related to the way the iterates of the
renormalization map approach certain curves (cf remark \ref{remarquedebut}). We hope
that this example will help to understand the general case of finitely
ramified self-similar sets.

The strategy we adopt here to obtain the renormalization equation
is a bit different than in \cite{Sabot5}, and much more straightforward:
we write a renormalization equation directly on the propagator of the
differential equation associed with the spectral problem. This simplification
comes from the 1-dimensional nature of the problem in this example.

\section{Definitions and results}

Let $I$ be the interval $I=[0,1]$ and $\alpha$ a real such that
$0<\alpha<1$. We set $\delta={\alpha\over 1-\alpha}$. We define
the two homotheties  $\Psi_1, \Psi_2$ by :
$$\Psi_1( x)=\alpha x,\;\;\;\Psi_2(x)= 1-(1-\alpha )(1-x).$$
So, $\Psi_1(I)=[0,\alpha]$, $\Psi_2(I)=[\alpha, 1]$ and the
interval $I$ is self-similar with respect to $(\Psi_1, \Psi_2)$.
\ali
 Let $b$ be a real number such that $0<b<1$. It is classical that
there exists a unique probability measure $m$ on $I$ such that
\begin{eqnarray} \label{f.1.1}
\int_0^1 fdm=b\int_0^1 f\circ \Psi_1 dm +
(1-b)\int_0^1 f\circ \Psi_2 dm,\;\;\; \forall f\in C(I).
\end{eqnarray}
N.B.: For $\alpha\ne b$, the measure $m$ is singular with respect
to the Lebesgue measure, for $\alpha=b$ it is the Lebesgue
measure.

 We denote by $H^+$ the operator ${d\over dm}{d\over dx}$ with
Neumann boundary conditions on $I$, i.e. it is the operator defined
on the domain:
\begin{eqnarray*}
\label{f.1.3.1} \{f\in L^2(I,m),\; \;  \exists g\in L^2(I,m),
f(x)=ax+b+
&\int _0^x \int _0^y g(z) dm(z) dy, \nonumber \\
&f'(0)=f'(1)=0\}, \nonumber
\\
\hbox{by\;\;\; $H^+f =g$.}
\end{eqnarray*}
We denote by $H^-$ the corresponding operator with Dirichlet
boundary conditions on $I$, i.e. the operator defined on the
domain
\begin{eqnarray*}
\label{f.1.3.1.1} \{f\in L^2(I,m),\; \;  \exists g\in L^2(I,m),
f(x)=ax+b+
&\int _0^x \int _0^y g(z) dm(z) dy, \nonumber \\
&f(0)=f(1)=0\}, \nonumber
\\
\hbox{by\;\;\; $H^-f =g$.}
\end{eqnarray*}
\Rm:
 These operators belong to the class of self-similar Laplacians as
 defined for example in \cite{Kigami1} or \cite{Sabot1}. Indeed,
it is clear that $H^+$ is the infinitesimal generator associated
with the classical Dirichlet form $a(f,g)=\int_0^1 f'g'dx$ defined
on $\ddd=\{f\in L^2(I,m), \; f'\in L^2(I,dx)\}$ and with the
measure $m$. By a change of variables we easily see that $a$
satisfies the following self-similarity relation
\begin{eqnarray*}
a(f,g)=\alpha^{-1}a(f\circ \Psi_1, g\circ \Psi_1)+
(1-\alpha)^{-1}a(f\circ\Psi_2, g\circ \Psi_2),\;\;\;\forall
f,g\in\ddd.
\end{eqnarray*}

Let $\Omega=\{1,2\}^\BN$ and let us fix a sequence
$\omega=(\omega_1,\ldots ,\omega_k,\ldots )$ in $\Omega$. We call this
sequence the blow-up. We extend the set $I$ to a
set $I_\nn (\omega)$ by scaling by
$$
I_\nn(\omega)=\Psi_{\omega_1}^{-1}\circ \cdots \circ
\Psi_{\omega_n}^{-1} (I).
$$
(We write $I_\nn(\omega)$ to show the dependence of $I_\nn$ in
$\omega$, but we will simply write $I_\nn$ when no ambiguity is
possible). Clearly, $I_\nn(\omega)\subset I_{<n+1>}(\omega)$ and
$\omega_{n+1}$ determines the position of $I_\nn(\omega)$ in
$I_{<n+1>}(\omega)$ (it is the left subinterval of
$I_{<n+1>}(\omega)$ when $\w_{n+1}=1$ and the right otherwise).
Then we set
$$
I_\infi(\w)=\cup_{n=0}^\infty I_\nn(\omega).
$$
We clearly see that $ I_\infi(\w)$ is either a half-line bounded
from the left if $\w$ is stationary to 1, a half-line bounded from
the right if $\omega$ is stationary to 2, or the real line if $\w$
is not stationary. We denote by $\partial I=\{0,1\}$ the boundary
points of $I$ and by $\partial I_\nn(\w)$, $\partial I_\infi(\w)$,
the boundary points of $I_\nn(\w)$, $I_\infi(\w)$. Of course,
$\partial I_\infi(\w)$ is empty when $\w$ is not stationary, and
contains a unique point when $\w$ is stationary.

We extend the measure $m$ to a measure $m_\nn$ on $I_\nn(\w)$ by
scaling by
\begin{eqnarray} \label{f.1.2}
\int_{I_\nn} f dm_\nn = b_{\w_1}^{-1} \cdots b_{\omega_n}^{-1}
\int_I f\circ \Psi_{\omega_1}^{-1}\circ \cdots \circ
\Psi_{\omega_n}^{-1} dm ,
\end{eqnarray}
where we set $b_1=b$, $b_2=1-b$. Clearly $b_\nn$ is a compatible
sequence of measures in the sense that if $\supp f\subset I_\nn$
then $ \int_{I_{<n+p>}} f dm_{<n+p>}=\int_{I_\nn} f dm_\nn$, for
all $p\ge 0$. Hence, it can be extended to a measure $m_\infi$
on $I_\infi$.

Then we define $H_\nn^+(\omega)$, $H_\nn^-(\w)$ (resp.
$H_\infi^+(\w)$, $H_\infi^-(\w)$) as the operators ${d\over
dm_\nn}{d\over dx}$ with Neumann or Dirichlet boundary conditions
on $I_\nn(\w)$ (resp. $I_\infi(\w)$). Of course, when $\partial
I_\infi(\w)=\emptyset$ then $H^+_\infi(\w)=H^-_\infi(\w)$ and we
simply write $H_\infi(\omega)$.

From now on, we make the following assumption
 \ali

(H) We choose $b=1-\alpha$, and we set
$\gamma=\alpha^{-1}(1-\alpha)^{-1}$.
 \ali \ali
 This choice for $b$ is crucial, it implies that the operator is
 locally invariant by translation, i.e. that in each subcell of
 $I_{<n>}$ of level $<p>$, $p\le n$ (i.e. in each subinterval of the
 type $\Psi_{\omega_1}^{-1}\circ \cdots \circ
\Psi_{\omega_n}^{-1}(\Psi_{j_1}\circ \cdots \circ \Psi_{j_p}(I)$)
the restriction of the operator $H_\nn$ is the same. This seems to
be the natural counterpart of the statistical invariance by
translation for random Schr\"odinger operators (cf \cite{Sabot5} for details).
 \ali

Thanks to the hypothesis (H), we can easily check that the
operators $H_\nn^\pm(\w)$ are isomorphic for different $\w$.
Indeed, if $\tilde \Psi$ is the right composition of $\Psi_i$ and
$\Psi_i^{-1}$ that sends $I_\nn(\omega)$ to $I_\nn(\w')$, then
$H_\nn^\pm(\w)(f\circ \tilde \Psi)=(H_\nn^\pm (\w')(f))\circ
\tilde \Psi$. But this is no longer true for $H_\infi(\w)$ (in
fact, $H_\infi(\w)$ and $H_\infi^\pm(\w')$ are isomorphic if $\w$
and $\w'$ are equal after a certain level, but this is a priori
not true otherwise, cf \cite{Sabot6} for precisions).

Let us denote by $\nu_\nn^+$ and $\nu^-_\nn$ the counting measures of
the spectrum of $H_\nn^+$ and $H_\nn^-$, respectively. By the
previous remark, $\nu_\nn^\pm$ does not depend on $\w$. We define
the density of states as the limit
$$
\mu =\lim_{n\to\infty } {1\over 2^n} \nu_\nn^\pm
$$
which exists and does not depend on the boundary condition $\pm$
(cf \cite{Fukushima1}, \cite{KigamiL} or \cite{Sabot5}).

We denote by $\Sigma^\pm(\w)$ the topological spectrum of
$H_\infi^\pm(\omega)$, and by $\Sigma^\pm_{ac}(\w)$,
$\Sigma^\pm_{sc}(\w)$, $\Sigma^\pm_{pp}(\w)$ respectively the
absolutely continuous, the singular continuous, the purely
ponctual part of the Lebesgue decomposition of the spectrum of
$H_\infi^\pm(\w)$. We also denote by $\Sigma_{ess}^\pm(\w)$ the
essential spectrum of $H_\infi^\pm(\w)$.
 We recall from \cite{Sabot6} the
following results (proved in the general setting of finitely
ramified self-similar sets).
\begin{propos}(\cite{Sabot6}, proposition 1) \label{p.1.1}
i) If the boundary set $\partial I_\infi(\w)$ is empty, i.e. if
$\w$ is not stationary, then
$\supp\mu=\Sigma(\w)=\Sigma_{ess}(\w)$.

ii) Otherwise we only have
$\supp\mu=\Sigma_{ess}^+(\w)=\Sigma_{ess}^-(\w)$. Moreover, the
eigenvalues eventually lying in $\Sigma^\pm(\w)\setminus
\supp(\mu)$ have multiplicity 1.
\end{propos}
\begin{propos}(\cite{Sabot6}, proposition 2) \label{p.1.2}
There exists deterministic sets $\Sigma, \Sigma_{ac}, \Sigma_{sc}
$ and $\Sigma_{pp}$ such that for almost all $\w$ in $\W$ (for the
product of the uniform measure on $\{1,2\}$) we have
$\Sigma_\bullet^\pm (\w)=\Sigma_\bullet$.
\end{propos}
In \cite{Sabot3}, we gave an explicit expression for the density
of states in terms of the Green function of a certain rational map
defined on the complex projective plane, and we proved that the
spectrum is a Cantor subset of $\BR_-$ for $\alpha\neq \demi$. The
aim of this text is to go further and to describe the spectral
type of the operator $H_\infi^\pm(\w)$. The main result is the
following
\begin{thm} \label{t.1.1}

i) If $I_\infi (\w)$ is a half-line bounded from the left, i.e. if
$\w$ is stationary to 1, then for the Neumann boundary conditions we
have
\begin{itemize}
\item
If $\delta>1$, the spectrum of $H_\infi^+$ is pure point and the
eigenvalues lie in the complement of $\supp\mu$.
\item
If $\delta<1$, then $\Sigma^+(\w)=\supp\mu$ and the spectrum of
$H_\infi^+(\w)$ is continuous, i.e. $\Sigma^+_{pp}(\w)=\emptyset$.
\end{itemize}
For the Dirichlet boundary condition we have
\begin{itemize}
\item
If $\delta>1$, then $\Sigma^-(\w)=\supp\mu$ and the spectrum of
$H_\infi^-(\w)$ is continuous, i.e. $\Sigma^-_{pp}(\w)=\emptyset$.
\item
If $\delta<1$, the spectrum of $H_\infi^-$ is pure point and the
eigenvalues lie in the complement of $\supp\mu$.
\end{itemize}

ii) If $I_\infi(\w)$ is the half-line bounded from the right, i.e.
if $\w$ is stationary to 2, then the same results hold, just
replacing $\delta$ by $\delta^{-1}$.

iii) If the boundary set $\partial I_\infi(\w)$ is empty, i.e. if the
blow-up $\w$ is not stationary, then
$\Sigma(\w)=\supp\mu$ and the spectrum of $H_\infi(\w)$ is
continuous, that is $\Sigma_{pp}(\w)=\emptyset$.
\end{thm}

\section{The renormalization equation of the propagator.
Expression of the density of states}

We suppose in this section that $\w=(1,\ldots ,1,\ldots)$ so that
$I_\infi(\w)=\BR_+$. We denote by $\Gamma_\lambda (s,t)$ the
propagator on $[s,t]\subset \BR_+$ of the equation
\begin{eqnarray*}
{d\over dm_\infi}{d\over dx} f=\lambda f,\hspace*{1.5in}
(E_\lambda)
\end{eqnarray*}
for $\lambda $ in $\BC$.
 This means that if $f$ is a solution of
$(E_\lambda)$ on $[s,t]$ then
$$
\left(
\begin{array}{c}
f(t)\\
f'(t)
\end{array}
\right) = \Gamma_\lambda(s,t)
\left(
\begin{array}{c}
f(s)\\
f'(s)
\end{array}
\right) .
$$
Classically, $\Gamma_\lambda (s,t)$ is an element of $Sl_2(\BC)$
and the coefficients of $\Gamma_\lambda (s,t)$ are analytic in
$\lambda$. We set $\Gamma_\lambda =\Gamma_\lambda(0,1)$ and
$\Gamma_{\nn,\lambda}=\Gamma_\lambda(0,\alpha^{-n})$.

We use the scaling invariance of the operator and the
self-similarity to deduce two fondamental relations. Let us
introduce the following notation: for a real $\beta$ we denote by
$D_\beta$ the $2\times 2$ matrix
$$
D_\beta= \left(
\begin{array}{cc}
1&0
\\
0&\beta \end{array} \right).
$$
The scaling relation of $m$, cf (\ref{f.1.2}), implies
\begin{eqnarray} \label{f.2.2}
\Gamma_{\nn,\lambda}=D_{\alpha^n}\circ \Gamma_{\gamma^n
\lambda}\circ D_{\alpha^{-n}},
\end{eqnarray}
where $\gamma=(\alpha(1-\alpha))^{-1}$, cf hypothesis (H). Indeed,
if $f$ is a solution of $(E_\lambda)$ on $I_\nn$, then $f\circ
\Psi_1^{-n}$ is a solution of $(E_{\gamma^n \lambda})$ on $I$, and
$$
\left(
\begin{array}{c}
f(\Psi_1^{-n}(t))\\
f'(\Psi_1^{-n}(t))
\end{array}
\right) = D_{\alpha^n} \left(
\begin{array}{c}
f\circ\Psi_1^{-n}(t)\\
(f\circ \Psi_1^{-n})'(t)
\end{array}
\right)
$$
for all $t$ in $I$.
 \ali
From the self-similarity, and hypothesis (H),  we can deduce
\begin{eqnarray}
\nonumber \Gamma_{<1>,\lambda} &=& \Gamma_\lambda
(1,\alpha^{-1})\circ \Gamma_\lambda(0,1)
\\
\label{f.2.3} &=& D_{\delta} \circ \Gamma_\lambda \circ
D_{\delta^{-1}}\circ \Gamma_\lambda,
\end{eqnarray}
where we recall that $\delta={\alpha\over 1-\alpha}$. Indeed, if
$f$ is solution on $I_{<1>}$ of $(E_\lambda)$, then $\tilde f=
f_{|[1,\alpha^{-1}]}\circ \Psi_1^{-1}\circ \Psi_2$ is solution of
$(E_\lambda)$ on $[0,1]$ and
$$
\left(
\begin{array}{c}
\tilde f(t)\\
\tilde f'(t)
\end{array}
\right) = D_{\delta^{-1}} \left(
\begin{array}{c}
f(\Psi^{-1}_1\circ \Psi_2(t))\\
f'(\Psi^{-1}_1\circ \Psi_2(t))
\end{array}
\right) , \;\;\;\forall t\in [0,1].
$$
We set
$$
\Gamma_{\nn,\lambda } = \left(
\begin{array}{cc}
a_\nn(\lambda) & b_\nn(\lambda )
\\
c_\nn(\lambda ) & d_\nn(\lambda )
\end{array}
\right)
$$
and simply $a_{<0>}=a$, $b_{<0>}=b$, $c_{<0>}=c$, $d_{<0>}=d$.

Using (\ref{f.2.2}) and (\ref{f.2.3})  an easy computation gives
\begin{eqnarray}
\nonumber
&&\left(
\begin{array}{cc}
a_{<1>}(\lambda) & b_{<1>}(\lambda )
\\
c_{<1>}(\lambda ) & d_{<1>}(\lambda )
\end{array}
\right)= \left(
\begin{array}{cc}
a(\gamma \lambda) & \alpha^{-1} b(\gamma \lambda )
\\ \label{f.2.4}
\alpha c(\gamma \lambda ) & d(\gamma \lambda )
\end{array}
\right)
\\
&=& \left(
\begin{array}{cc}
a(\lambda)(a(\lambda )+\delta^{-1} d(\lambda))-\delta^{-1} &
b(\lambda)(a(\lambda )+\delta^{-1} d(\lambda))
\\
\delta c(\lambda )(a(\lambda )+\delta^{-1} d(\lambda))
 & \delta d(\lambda )(a(\lambda )+\delta^{-1} d(\lambda))-\delta
\end{array}
\right)
\end{eqnarray}
We introduce the polynomial map
\begin{eqnarray} \nonumber
f:&\BC^2&\rightarrow \BC^2
\\
\label{f.2.4.0}
 &(x,y)&\rightarrow  (x(x+\delta^{-1}
y)-\delta^{-1}, \delta y(x+\delta^{-1}y)-\delta )
\end{eqnarray}
and we see from (\ref{f.2.4}) that the map $\phi:\BC\rightarrow
\BC^2$ given by
$$
\phi(\lambda)=\left(
\begin{array}{c}
a(\lambda) \\
d(\lambda)
\end{array}
\right)
$$
satisfies
\begin{eqnarray} \label{f.2.5}
f\circ \phi(\lambda) =\phi(\gamma \lambda),\;\;\;\forall \lambda
\in \BC.
\end{eqnarray}
Hence, $\{\phi(\lambda),\lambda\in \BC\}$ is an invariant
complex curve for $f$.
\begin{propos}\label{p.2.1}
(i) The Neumann and Dirichlet spectrum of ${d\over dm_\nn}{d\over
dx}$ are given by
\begin{eqnarray*}
\nu_\nn^+&=&{1\over 2\pi} \Delta \ln\vert c_\nn(\lambda )\vert ,
\\
\nu_\nn^-&=&{1\over 2\pi} \Delta \ln\vert b_\nn(\lambda )\vert ,
\end{eqnarray*}
where $\Delta$ denotes the distributional Laplacian on $\BC$.

(ii) The infinite product
$$ \prod_{k=1}^\infty
\alpha( a(\gamma^{-k}\lambda)+{\delta}^{-1} d(\gamma^{-k}\lambda))
$$
is convergent and we have
\begin{eqnarray*}
&b(\lambda)=\prod_{k=1}^\infty \alpha(
a(\gamma^{-k}\lambda)+{\delta}^{-1} d(\gamma^{-k}\lambda)),
\\
&c(\lambda)=\lambda \prod_{k=1}^\infty \alpha(
a(\gamma^{-k}\lambda)+{\delta}^{-1} d(\gamma^{-k}\lambda)).
\end{eqnarray*}
\end{propos}
Proof: (i) The eigenvalues of $H_\nn^+$ and $H_\nn^-$ are simple
and it is clear that the sets of zeroes of $c_\nn$ and $b_\nn$
coincide respectively with the sets of eigenvalues of $H_\nn^+$
and $H_\nn^-$. Thus, the only thing to prove is that the zeroes of
$c_\nn$ and the zeroes of $b_\nn$ are simple. By scaling it is
enough to prove it for $c$ and $b$. Consider a complex number
$\lambda$ such that $\im \lambda >0$ and denote by $f$ the
solution of $(E_\lambda)$ on $I$ with initial condition $f(0)=1$,
$f'(0)=0$. By the Lagrange identity we have
$$
[\im(\overline f f')]_0^1=\im(\lambda) \int_I \vert f\vert^2 dm,
$$
which in our case gives
\begin{eqnarray}\label{f.2.5.1}
\im(\overline a(\lambda)c(\lambda))=\im(\lambda)\int_I\vert
f\vert^2dm.
\end{eqnarray}
In particular, this means that $\overline a(\lambda)c(\lambda)$
sends the upper-half plane to itself and hence $\overline
a(\lambda)c(\lambda)$ cannot have a multiple zero on the real
axis.

(ii) Remark first that $a(0)=b(0)=d(0)=1$ and $c(0)=0$. Hence,
there exists $K$ such that $c(\lambda)=K\lambda +O(\lambda^2)$ for
$\lambda$ small. Considering equation (\ref{f.2.5.1}) for
$\lambda$ small, we deduce that $K=1$. Indeed, $\overline a (0)=1$
and thus we have $K=\int_I \vert f\vert^2 dm$, where $f$ is the
solution of ${d\over dm}{d\over dx} f=0$, with initial condition
$f(0)=1$, $f'(0)=0$. But this solution is $f=1$.

From relation (\ref{f.2.4}) we deduce
\begin{eqnarray*}
b(\lambda)&=& b(\gamma^{-n}\lambda) \prod_{k=1}^n \alpha(
a(\gamma^{-k}\lambda)+{\delta}^{-1} d(\gamma^{-k}\lambda))
\\
&{\sim\atop n\to \infty}& \prod_{k=1}^n \alpha(
a(\gamma^{-k}\lambda)+{\delta}^{-1} d(\gamma^{-k}\lambda))
\end{eqnarray*}
and
 \begin{eqnarray*}
 c(\lambda)&=& c(\gamma^{-n} \lambda)\prod_{k=1}^n
(1-\alpha)^{-1}( a(\gamma^{-k}\lambda)+\sqrt{\delta}^{-1}
d(\gamma^{-k}\lambda))
\\
&{\sim \atop n\to \infty}& \lambda  \prod_{k=1}^n \alpha^{-1}(
a(\gamma^{-k}\lambda)+{\delta}^{-1} d(\gamma^{-k}\lambda))
\end{eqnarray*}
This immediately implies that the product is convergent and the
formulas for $b(\lambda)$ and $c(\lambda)$.$\diamondsuit$

\subsection{ The Green function of $f$. Elements of the dynamics
of $f$.}

The map $f$ has a natural extension to the 2-dimensional
projective space $\BP^2$, given in homogeneous coordinates by
\begin{eqnarray} \label{f.2.6}
f([x,y,z]) = [x(x+\delta^{-1} y)-\delta^{-1}z^2, \delta
y(x+\delta^{-1}y)-\delta z^2, z^2].
\end{eqnarray}
(The point $[x,y,z]$ represents the image of $(x,y,z)$ in $\BC^3$
by the canonical projection $\pi:\BC^3\rightarrow \BP^2$).
Equivalently, this means that $f$ is the map induced on $\BP^2$ by
the homogeneous polynomial map $R:\BC^3\rightarrow \BC^3$ given by
$$
R((x,y,z)) = (x(x+\delta^{-1} y)-\delta^{-1}z^2, \delta
y(x+\delta^{-1}y)-\delta z^2, z^2),
$$
the relation being $f(\pi(x))=\pi(R(x))$. We can see that $f$ is
not defined where $R$ is null. We remark that $R((1,-\delta,
0))=(0,0,0)$ and that ${\Bbb C}(1,-\delta,0)$ is the unique
complex line on which $R$ is null. We denote by $l=[1,-\delta ,0]$
the associated point in the projective space and we say that $l$
is a point of indeterminacy. The map $f$ is then a map from
$\Pp\setminus\{l\}$
 to $\Pp$ and is holomorphic on $\Pp \setminus \{l\}$
(in fact the image of the point $l$ by $f$ can be defined as a
compact Riemann surface called the blow-up of $l$, cf
(\cite{Sabot3}). Therefore, the map $f$ is called a rational map of
$\Pp$. Its degree is 2 in relation with the degree of the
homogeneous polynomials appearing in $R$ (cf \cite{Sibony1}).
 \ali We set
\begin{eqnarray}
\label{f.2.7} D=\{[x,y,z],\;\;x+\delta^{-1}y=0\}.
\end{eqnarray}
 The line $D$ is sent by $f$ to a unique point,
$[-\delta^{-1},-\delta,1]$, and the orbit of $D$ is
\begin{eqnarray} \label{f.2.8}
f(D\setminus\{l\})=\{[-\delta^{-1},-\delta,1]\},\;\hbox{ and }\;
f^{n+1}(D\setminus \{l\})=[\delta^{-2^n}, \delta^{2^n}, 1],
\end{eqnarray}
for $n\ge 0$. The set $D$ is called a $f$-constant curve (line), since it
is sent to a unique point.
It is a general phenomenon that a $f$-constant curve contains a
point of indeterminacy (cf proposition 1.2 of \cite{Sibony1}).
 \ali Another
important property of the map $f$ is that it has no degree
lowering curve: a degree lowering curve is a f-constant curve sent
by $f^n$ to a point of indeterminacy. When such a phenomenon
appears, a common factor, which can be divided out, appears in
$R^n$ and the degree of the map $f^n$ drops. Here, we remark that
the orbit of $D$ does not contain the indeterminacy point $l$, so
$D$ is not a degree lowering curve. Hence, $\hbox{degree}
(f^n)=2^n$ (this means that $R^n$ is represented by 3 homogeneous
polynomials of degree $2^n$ with no common factor).
 Following the terminology in \cite{Sibony1}, $f$ is said
to be algebraically stable, cf definition 4.4.
 \ali We set $\tilde
I=\cup_{n\ge 0}f^{-n}(\{l\})= \{[1,-\delta^{-(n-1)},0]\}_{n\ge 0}$
the set of preimages of the point of indeterminacy $\{l\}$. \ali
The Fatou set of $f$ is defined to be the union of all open balls
$U\subset \Pp\setminus \tilde I$ on which the family
$\{f^n\}_{n\ge 0}$ is normal. The Fatou set is denoted by $\fff$
and its complement, the Julia set, by $\jjj=\Pp\setminus\fff$. Of
particular interest to us is the fact that the attractive basin of
an attractive fixed point is in the Fatou set.
 \ali A
function, useful to study the dynamics of $f$, is the Green
function defined as the limit of the sequence of functions
$G_n:\;{\Bbb C}^2\rightarrow {\Bbb R}\cup\{-\infty\}$:
\begin{eqnarray} \label{f.2.9}
G_n(x)={1\over 2^n}\log (1+\| f^n(x)\|),\;\;\;x\in {\Bbb C}^2.
\end{eqnarray}
where $\|\;\|$ denotes the usual norm of ${\Bbb C}^2$, and where
we considered $f$ as a map on $\BC^2$, by (\ref{f.2.4.0}). \ali We
will use the following result (cf \cite{Sibony1}, proposition 2.11
or \cite{FSibony}, theorem 1.6.1):
\begin{propos} \label{p.2.2}
(i) The limit
$$
G (x)=\lim_{n\to \infty} G_n(x)
$$
exists for all $x\in {\Bbb C}^2$. The function $G$ is
plurisubharmonic and satisfies
\begin{eqnarray}
\label{f.3.1.0} G\circ f=2 G.
\end{eqnarray}
 (ii) $G$ is pluriharmonic on $\BC^2\cap \fff$.
\end{propos}
NB: In general for a rational map on $\BP^k$ the Green function is
defined on $\BC^{k+1}$, cf \cite{Sibony1}. In our case since $f$
is polynomial it is easier to adopt the previous definition.

Denote by $\zeta(\lambda )$ the Lyapounov exponent of the
propagator of the O.D.E. $(E_\lambda)$ given by
$$
\zeta (\lambda) =\lim_{n\to \infty} {1\over 2^n} \ln\|
\Gamma_\lambda (0,\alpha^{-n})\|
$$
when this limit exists. In \cite{Sabot3}, theorem 3.1 
and proposition 3.6,  we proved
\begin{thm}\label{t.2.1}
(i) The Lyapounov exponent exists and is given by
$$
\zeta(\lambda)=G\circ \phi(\lambda)
$$
for all $\lambda$ in $\BC$.

(ii) The density of states is given by
$$\mu={1\over 2\pi} \Delta
\zeta,
$$
where $\Delta$ denotes the distributional Laplacian on $\BC$.
\end{thm}
Rm: The point (ii) corresponds to the Thouless formula in our
context. In \cite{Sabot5} we gave similar formulas in the general
case of finitely ramified self-similar sets for the density of
states in terms of the Green function of a certain renormalization
map.

In \cite{Sabot3} we deduced from the analysis of the dynamics of
$f$ that $\supp(\mu)$ is a Cantor subset of $\BR_-$ for
$\delta\neq \demi$, and for certain values of the parameter
$\delta$ we proved the local H\"older regularity of the Lyapounov
exponent.

\subsection{Some details about the dynamics of the map $f$}
We give some information about the dynamics of the map $f$ that
will be useful in the sequel.

We set $x_+=[0,1,0]$ and $x_-=[1,0,0]$. The first important point
is that for $\delta>1$
\begin{itemize}
\item
$x_-$ have one attractive direction and one repulsive (with
eigenvalues 0 and $\delta$).
\item
$x_+$ is attractive (with eigenvalues 0 and $\delta^{-1}$).
\end{itemize}
(and the reverse for $\delta<1$).

We denote by $\ccc\subset\Pp$ the hypersurface:
\begin{eqnarray}
\label{f.2.1.1} \ccc=\{[x,y,z],\;\;xy=z^2\}.
\end{eqnarray}
The restriction of $\ccc$ to $\BR^2$ consists of two branches of
hyperbolas (cf picture 1). Remark that the restriction of $f$ to
$\ccc$ is given by
$$
f([x,y,z])=[x^2,y^2,z^2], \;\;\; [x,y,z]\in \ccc,
$$
and thus
$$
f(C)\subset C.
$$
We set
\begin{eqnarray} \label{f.2.1.2}
\ccc_+=\{[x,y,z]\in \ccc,\; \vert x\vert <\vert y\vert \},\;\;\;
\ccc_-=\{[x,y,z]\in \ccc,\; \vert x\vert >\vert y\vert \}.
\end{eqnarray}
Clearly, any point in $\ccc_\pm$ converges to $x_\pm$. Hence, the
set $\ccc_+$ is in the Fatou set of $f$ for $\delta>1$ since it is
in the attractive basin of $x_+$. The same is true for the set
$D\setminus \{l\}$ since $f^n(D\setminus \{l\})$ converges to
$x_+$ (cf formula (\ref{f.2.8}).

We set
\begin{eqnarray}
\label{f.2.1.3} K_+\;\; (\hbox{resp. $K_-$})=
\{[x,y,z],\;(x,y,z)\in\BR^3,\; x y-z^2\ge 0\}\;\;(\hbox{resp. $x
y-z^2 \le 0$}).
\end{eqnarray}
The following properties are direct consequences of the
forthcoming formula \ref{f.3.n.0} (cf
\cite{Sabot3}, formula 3.17)
\begin{eqnarray}
\label{f.2.1.4} f(K_\pm)\subset K_\pm,\;\;\; \phi(\BR_\pm)\subset
K_\pm.
\end{eqnarray}

\section{Spectral analysis of the operator on a half-line}
In this section, we consider the case where $\w$ is stationary, i.e.
where $\partial I_\infi (\w)\neq \emptyset$.
By the symmetric role played by $\alpha$ and $(1-\alpha)$ it is
enough to analyse the case where $I_\infi(\w)$ is a half-line bounded
from the left, i.e. when $\w$ is stationary to $1$. By scaling it
is enough to consider the case $\w=(1,\ldots ,1,\ldots )$. Thus, in this
section, we only consider the case $\w=(1,\ldots ,1,\ldots )$, so that
$I_\infi (\w) =\BR_+$.

Denote by $S$ the set of intersection times of the curve
$\phi(\gamma^{-1}\lambda)$ with the line $D$ defined in
(\ref{f.2.7}), i.e.
$$
S=\{\lambda, \;\; \phi(\gamma^{-1} \lambda )\in D\}.
$$
 For $p$ in $\BZ$ we set
$$
S_p=\gamma^{p} S.
$$
We know from proposition \ref{p.2.1}  that the spectrum of
$H_{<0>}^+$ and $H_{<0>}^-$ are given by
\begin{eqnarray}\label{f.3.0.0}
\supp\nu_{<0>}^+\setminus\{0\}=\supp\nu_{<0>}^- =
\cup_{p=0}^\infty S_p.
\end{eqnarray}
Hence, we know that $S$ is non-empty and included in $\BR_-$. The
set $S$ is also infinite. Indeed, if $\lambda_0$ is in $S$ then
$\phi(\gamma^{-1}\lambda_0)$ is in $D$ and
$\phi(\lambda_0)=[-\delta^{-1},-\delta,1]$,
$\phi(\gamma\lambda_0)=[\delta^{-2},\delta^{2},1]$, cf
(\ref{f.2.8}). Hence, $\phi(\gamma^{-1}\lambda)$ must necessarily
cross the line $D$ between $\gamma \lambda_0$ and $\gamma^2
\lambda_0$.  We denote by $0>\lambda_1>\lambda_2> \ldots
>\lambda_k>\ldots$ the ordered set of points of $S$. We also
denote by $\lambda_{k,p}=\gamma^p\lambda_k$ the points of $S_p$.

For any $k>0$ we denote by $f_k^\pm$ the solutions of the equation
$(E_{\lambda_k})$ with initial condition $f^+_{k}(0)=1$,
$(f^+_{k})'(0)=0$ and $f^-_{k}(0)=0$, $(f^-_{k})'(0)=1$. We denote
by $f_{k,p}^\pm$, for $p$ in $\BZ$, the scaled copy of $f^\pm_k$
$$
f_{k,p}^\pm=f_k^\pm\circ \Psi_1^{-p}.
$$
By scaling $f_{k,p}^\pm$ is solution of $(E_{\gamma^p\lambda_k})$
on $\BR_+$. We now denote by $f_{k,p,\nn}^\pm$ the restriction of
$f_{k,p}^\pm $ to $I_\nn$.
\begin{lem}
(i)
The set
$$
\{1,\; f^+_{k,p,\nn},\; k>0,\; p\ge -n\},
$$
is a complete set of eigenfunctions of $H_\nn^+$.

(ii) The set
$$
\{f_{k,p,\nn}^-, \; k>0, \; p\ge -n\}
$$
is a complete set of eigenfunctions of $H_\nn^-$.
\end{lem}
Proof: (i) By scaling it is enough to check it for $n=0$. We
already know that the eigenvalues of $H_{<0>}^+$ are equal to the
set $\{0\}\cup_{p=0}^\infty S_p$. The functions $f_{k,p,<0>}^+$
satisfy the Neumann boundary condition in 0 and are solutions of
$(E_{\gamma^p\lambda})$ on $I$, hence $f_{k,p,<0>}$ is necessarily
an eigenfunction of $H_{<0>}^+$ on $I$ for $p\ge 0$. Since the
eigenvalues of $H_{<0>}^+$ are simple, $1$ and $f_{k,p,<0>}$ for
$p\ge 0$, $k\ge 1$,  do necessarily form a complete family of eigenfunctions of
$H_{<0>}^+$. The proof of (ii) is similar.$\diamondsuit$
 \ali

\begin{thm}\label{t.3.1}
 (The pure point case)

(i) If $\delta>1$, the functions $\{f_{k,p}^+\}_{k\ge1,p\in\BZ}$
are in $L^2(\BR_+, m_\infi)$ and form a complete family of
eigenfunctions of $H^+_\infi$. Hence, the spectrum of $H_\infi^+$
is pure point and the set of eigenvalues is
\begin{eqnarray} \label{f.3.spect}
\cup_{p=-\infty}^\infty S_p.
\end{eqnarray}
These eigenvalues lie  in the complement of $\supp\mu$.

(ii) If $\delta<1$, the functions $\{f_{k,p}^-\}_{k>0,p\in\BZ}$
are in $L^2(\BR_+, m_\infi)$ and form a complete set of
eigenfunctions of $H_\infi^-$. Hence the spectrum of $H_\infi^-$
is purely ponctual and the set of eigenvalues is
(\ref{f.3.spect}).
\end{thm}
Proof: (i) By scaling it is enough to prove that $f_{k,p}^+$ is in
$L^2$ for $p=0$. Considering the orbit of the line $D$, cf
(\ref{f.2.8}), we deduce from (\ref{f.2.2}) and (\ref{f.2.4}) that
for a point $\lambda_k$ in $S$ we have
$$
\Gamma_{<0>,\lambda_k} = \left(
\begin{array}{cc}
-\delta^{-1} & 0
\\
0& -\delta
\end{array}
\right)  \; \hbox{ and } \;
 \Gamma_{<n>,\lambda_k} = \left(
\begin{array}{cc}
\delta^{2^{-n}} & 0
\\
0& \delta^{2^n}
\end{array}
\right)
$$
for $n\ge 1$.
 Thus we have
$$
\left(
\begin{array}{c}
f^+_k(1)
\\
(f^+_k)'(1)
\end{array}
\right) = \left(
\begin{array}{c}
-\delta^{-1}
\\
0
\end{array}
\right) \; \hbox{ and }\; \left(
\begin{array}{c}
f^+_k(\alpha^{-n})
\\
(f^+_k)'(\alpha^{-n})
\end{array}
\right) = \left(
\begin{array}{c}
\delta^{-2^n}
\\
0
\end{array}
\right)
$$
for $n\ge 1$.
 Hence for $n\ge 0$ the function $f^+_{k,<n+1>}$ can be
written
\begin{eqnarray}\label{f.3.0.1.0}
\left\{
\begin{array}{l}
(f^+_{k,<n+1>})_{|I_{<n>}}=f^+_{k,<n>}
\\
(f^+_{k,<n+1>})_{|I_{<n+1>}\setminus I_{<n>}}=b_n^+ \left(
f^+_{k,<n>}\circ \Psi_1^{-n} \circ \Psi_2^{-1}\circ \Psi_1^{n+1}
\right)
\end{array}
\right .
\end{eqnarray}
where $b_0^+=-\delta^{-1}$ and $b_n^+=\delta^{-2^n}$ for $n\ge 1$.
Indeed, the function on the right hand side of the second line is
a solution of the equation $(E_{\lambda_k})$ on
$I_{<n+1>}\setminus I_{<n>}$ and it matches exactly the boundary
condition of $f^+_{k,<n>}$. From this we deduce the following
$$
\| f_{k,<n+1>}^+\|^2 =\| f_{k,<n>}^+\|^2(1+\delta (b_\nn^+)^2),
$$
where $\| \;\|$ is the $L^2$ norm with respect to the measure
$m_{<n>}$ and $m_{<n+1>}$ (the extra factor $\delta$ comes from
the scaling relation between $m_\nn$ and
$(m_{<n+1>})_{|I_{<n+1>}\setminus I_\nn}$). For $\delta
>1$, this immediately implies that $f_k^+$ is in $L^2(m_\infi)$
for all $k>1$. Remark also that by scaling we deduce the following
relation, for all $n\ge 0$ and $p\ge -n$:
$$
\| f_{k,p,<n+1>}^+\|^2 =\| f_{k,p,<n>}^+\|^2(1+\delta
(b_{n+p}^+)^2),
$$
In particular, we deduce from the previous relation that there
exists a constant depending only on $\delta$,
$C_\delta=\prod_{k=0}^\infty (1+\delta (b_k^+)^2)$, such that for
all $p\ge -n$
\begin{eqnarray} \label{f.3.0.2}
\| f_{k,p}^+\|^2 \le C_\delta \| f_{k,p,<n>}\|^2.
\end{eqnarray}

To prove that the family is complete it is enough to prove that
for any function $g$ in $L^2(m_\infi)$ with compact support we
have
\begin{eqnarray}\label{f.3.0.2.1}
\|g\|^2 =\sum_{k=1}^\infty \sum_{p\in \BZ} {\vert\int g f^+_{k,p}
dm_\infi \vert^2\over \| f_{k,p}^+\|^2 },
\end{eqnarray}
and we may as well suppose that $\supp g\subset I$ by scaling
invariance.
 \ali
Take a positive $\epsilon$. Since the sum on the right is
converging, we can find $n_1>0$ such that
\begin{eqnarray} \label{f.3.0.3}
\sum_{k=1}^\infty \sum^{-n_1-1}_{p=-\infty} {\vert \int g
f^+_{k,p} dm_\infi \vert^2\over \| f_{k,p}^+\|^2 } \le \epsilon .
\end{eqnarray}
 Since
$\{1, f_{k,p,\nn}^+, k\ge 1, p\ge -n\}$ is a complete family of
eigenfunctions for $H_\nn^+$ and since $\supp g\subset I$, we
have:
$$
\| g\|^2={\vert\int_{I_\nn} gdm_\nn\vert^2\over \int_{I_\nn } 1
dm_\nn}+ \sum_{k=1}^\infty \sum_{p=-n}^\infty {\vert \int_{I_\nn}
g f_{f,p,\nn}^+ dm_\nn\vert^2 \over  \| f_{k,p,\nn}^+\|^2}.
$$
For $n\ge n_1$ we have
\begin{eqnarray*}
&& \left\vert \| g\|^2-\sum_{k=1}^\infty \sum_{p=-n_1}^\infty
{\vert \int_{I_\infi} g f_{k,p}^+ dm_\infi \vert^2 \over  \|
f_{k,p,\nn}^+\|^2}\right\vert
\\
 &\le&
 {\vert\int_{I}
gdm \vert^2\over \int_{I_\nn } 1 dm_\nn}+ \sum_{k=1}^\infty
\sum_{p=-n}^{-n_1-1} {\vert \int_{I_\infi} g f_{k,p}^+ dm_\infi
\vert^2 \over  \| f_{k,p,\nn}^+\|^2}
\\
&\le&
 {\vert \int_I gdm\vert^2 \over (1-\alpha)^{-n}} + \epsilon C_\delta.
\end{eqnarray*}
In the last equation we used relation (\ref{f.3.0.2}) and
(\ref{f.3.0.3}). Letting $n$ go to infinity we get
$$
\left\vert \| g\|^2-\sum_{k=1}^\infty \sum_{p=-n_1}^\infty {\vert
\int_{I_\infi} g f_{k,p}^+ dm_\infi \vert^2 \over  \|
f_{k,p}^+\|^2}\right\vert \le \epsilon C_\delta.
$$
Letting $\epsilon $ go to zero we prove relation
(\ref{f.3.0.2.1}).

To prove that $\cup_{p\in \BZ} S_p$ is in the complement of
$\supp\mu$, we simply use the fact that the line $\BC^2\cap D$ is
in the attractive bassin of the point $[0,1,0]$ (which is
attractive for $\delta>1$, cf \cite{Sabot3}), and hence is in the
Fatou set of $f$. Thus, $\mu$ must be null in a neighborhood of
$\cup_{p\in \BZ} S_p$ since $G$ is pluriharmonic in a neighborhood
of $\BC^2\cap D$.

(ii) The proof for the Dirichlet boundary condition is similar. We
prove a similar relation on the norm of the functions
$f^-_{k,p,\nn}$ involving a sequence of coefficients $b_n^-$ given
by $b_0^-=-1$, $b_n^-=\delta^{2^n-1}$. Then the proof goes exactly
the same way.$\diamondsuit$

\begin{thm}\label{t.3.2}
 (The continuous case)

(i) If $\delta <1$, then in the case of the operator with Neumann
boundary conditions $\Sigma^+=\supp \mu$, and the spectrum of
$H_\infi^+$ is continuous.

(ii) If $\delta >1$ then in the case of the operator with
Dirichlet boundary condition $\Sigma^-=\supp \mu$, and the
spectrum of $H_\infi^-$ is continuous.
\end{thm}
Proof: We first prove the following lemma.
\begin{lem}
\label{l.3.0.2} For a stationary blow-up $\w$, we have
$$\Sigma^\pm \subset \cup_{p\in \BZ}S_p\sqcup \supp\mu.$$
\end{lem}
\Rm: This result is actually true for any blow-up $\w$, but for a
non-stationary blow-up $\w$ we proved in proposition \ref{p.1.1}, that
$\Sigma(\w)=\supp \mu$. Comparing this lemma with proposition \ref{p.1.1},
we see that it can be rephrased as
$\Sigma^\pm\setminus\Sigma^\pm_{ess}\subset \cup_{p\in \BZ} S_p$.
 \ali
Proof: From general results and formula (\ref{f.3.0.0}) we know
that
\begin{eqnarray*}
\Sigma^\pm &\subset& \cap_{n\in\BN} \overline{\cup_{m\ge n}\supp
\nu_{<m>}^\pm},
\\
&=&\overline{\cup_{p\in \BZ} S_p}
\end{eqnarray*}
Let us first prove that the points of $S_p$ are isolated in $\cup_{p\in
\BZ} S_p$. By scaling it is enough to prove it for $p=0$.  For
$\lambda_{k}\in S$, $\phi(\gamma^{-1}\lambda_{k})$ is in $D$ and
thus in the attractive basin of $x_+$ for $\delta> 1$ (and $x_-$
for $\delta<1$). We can find a neighborhood $U$ of $\lambda_{k}$
such that $f^n(\phi(\gamma^{-1}U))\cap D=\emptyset$ for all $n>0$.
This implies that $U\cap (\cup_{p< 0} S_p)=\emptyset$. To see that
$\lambda_k$ is isolated in $\cup_{p>0} S_p$ it is enough to remark
that $\cup_{p> 0} S_p$ has no accumulation point since for any
$R>0$ there exists $N$ large enough such that $S_p\cap
B(0,R)=\emptyset$ for $p\ge N$ ($B(0,R)$ is the ball in $\BC$ with
center $0$ and radius $R$). From this we deduce
$$
\Sigma^\pm \subset (\cup_{p\in \BZ} S_p)\sqcup (\cap_{n\in \BN}
\overline{\cup_{m\le -n} S_m}).
$$
Take now $\lambda$ in $\cap_{n\in \BN} \overline{\cup_{m\le -n}
S_m}$. Suppose that $\lambda $ is not in $\supp\mu$. This means
that there exists a small open ball $U$ around $\lambda$ such that
$\mu(U)=0$, thus that $G\circ \phi$ is harmonic on $U$. Mimicking
the proof of \cite{Sibony1}, Theorem 6.5, we see 
that this implies that the family of
functions $(f^n\circ \phi)_{n\in \BN}$ is normal on $U$. But
necessarily $U$ contains a point of $S_p$ for a certain $p<0$,
which is in the attractive basin of $x_+$ for $\delta>1$ (or $x_-$
for $\delta<1$). This implies that $U$ itself is included in the
attractive basin of $x_+$ or $x_-$. Considering a neighborhood $V$
of $x_+$ (or $x_-$) such that $V\cap D=\emptyset$, we know that
there exists $N$ such that $f^n(U)\subset V$ for $n\ge N$. Thus
$U\cap S_p=\emptyset$ for $p\le -(N+1)$ and this is
contradictory.$\diamondsuit$

To prove i) of theorem \ref{t.3.2} we first prove that for
$\delta<1$, $\lambda_{k,p}$ cannot be an eigenvalue of
$H_\infi^+$. But this is clear from relation (\ref{f.3.0.1.0})
since $f^+_{k,p}$ is not in $L^2(m_\infi)$. The only thing which remains
to prove is that
that $\lambda$ in $\supp\mu$ cannot be an
eigenvalue of $H_\infi^\pm$. This will be done in lemma 
\ref{l.noteigenvalue} in the next section.

The proof of ii) is strictly similar.

\section{The continuous spectrum}
We prove the following lemma.
\begin{lem} \label{l.noteigenvalue}
For any blow-up $\w$, $\lambda$ in $\supp \mu$ cannot be an eigenvalue
of  $H_\infi^\pm (\w)$.
\end{lem}
As we explained at the end of the previous section,
this lemma concludes the proof of 
Theorem \ref{t.3.2}.
Thanks to proposition \ref{p.1.1}, it also concludes the proof
of theorem \ref{t.1.1}, iii) (the case of a non-stationary blow-up
$\w$).
 \ali
We first prove
\begin{lem} \label{l.3.1}
For any $\lambda$ in $\supp\mu$, there exists a constant $C_1$,
depending only on $\lambda$ and $\delta$ such that for all $n$ in
$\BN$
$$ \| f^n(\phi(\lambda))\| \le C_1.$$
\end{lem}
\Rm: \label{remarquedebut} Let us remark that this implies that the iterates of
$\phi(\lambda)$ cannot approach the indeterminacy point of
$f$, which is located on the curve at infinity $[x,y,0]$.
It seems intuitively natural that this condition is related
to continuous spectrum since indeterminacy points correspond to 
eigenfunctions with compact support (cf \cite{Sabot5}). Hence,
eigenfunctions with non-compact support could be related to the
way an orbit can approach the set of indeterminacy points.
 \ali
Proof: If $\lambda=0$, then $\phi(\lambda)$ is fixed by $f$.
Let us take $\lambda $ in $\supp\mu\setminus \{0\}$. We set
$(x_0,y_0)=\phi(\lambda)$. The vector $(x_0,y_0)$ is in $K_-$ (cf
section 2.2). The result of lemma \ref{l.3.1} concerns only the
real dynamics of $f:\BP^2_\BR\rightarrow \BP_\BR^2$, where
$\BP^2_\BR$ is the 2-dimensional real projective space.
 \ali
We set for any $(x,y,z)$ in $\BR^3$, as in \cite{Sabot3},
\begin{eqnarray*}
&&r((x,y,z))=xy-z^2
\\
&&p((x,y,z))=\alpha(x+\delta^{-1}y).
\end{eqnarray*}
For any $\epsilon >0$, we set
$$V_\epsilon=\{[x,y,z]\in \BP_\BR^2, \;\; \vert r({(x,y,z)\over
\|(x,y,z)\|})\vert \le\epsilon\}.
$$
The family $(V_\epsilon)_{\epsilon>0}$ gives a base of
neighborhoods of $C\cap\BP_\BR^2$.
 We first prove the following lemma
\begin{lem}
There exists $\epsilon >0$ such that the iterates $f^n((x_0,y_0))$ do not enter
$V_\epsilon$, i.e.
$$
(\cup_{n\ge 0} \{f^n((x_0,y_0))\})\cap V_\epsilon =\emptyset.
$$
\end{lem}
Proof:
The set $C_+$ is in the attractive basin of $x_+$. Let $V_+$ be a
neighborhood in $\BP^2_\BR$ of $(C_+\cup\{x_+\})\cap\BP^2_\BR$
contained in the attractive basin of $x_+$. Since $(x_0,y_0)$ is
in the Julia set of $f$, the iterates $f^n((x_0,y_0))$ do not
enter the set $V_+$.
Remark that the closure in
$\BP_\BR^2$ of $C_-\cap\BP_\BR^2$ is given by
$$
\overline{C_-\cap\BP_\BR^2}=\{x_-\}\cup\{[x,{1\over x},1], \;\;
x\in \BR_*, \;\; \vert x\vert \ge 1\}.
$$
We now use, as in \cite{Sabot3}, (3.27), the following formula
\begin{eqnarray}
\label{f.3.n.0}
 && r\circ R(X) ={1\over \alpha(1-\alpha)} (p(X))^2
r(X), \;\;\; \forall X\in \BR^3,
\end{eqnarray}
which implies
\begin{eqnarray}
\label{f.3.n.1} \left\vert r\left({R(X)\over \| R(X)\|}\right)\right\vert =
 {\vert r\circ R(X)\vert \over
\| R(X)\|^2}={1\over \alpha(1-\alpha)}{\|X\|^2\vert
p(X)\vert^2\over \|R(X)\|^2} \left\vert r\left({X\over \|X\|}\right)\right\vert^2,
\;\;\; \forall X\in \BR^3.
\end{eqnarray}
Take $X=(x,y,z)$ s.t. $\pi(X)\in \overline{C_-\cap\BR^2}$. We have
\begin{eqnarray*}
{\|X\|^2\over \|R(X)\|^2} {1\over \alpha(1-\alpha)} \vert
p(X)\vert^2 &=& {x^2+y^2+z^2\over x^4+y^4+z^4} \delta
(x+\delta^{-1}y)^2.
\end{eqnarray*}
If $\pi(X)=x_-$ then the last value equals $\delta$. Otherwise, we
may as well suppose, by homogeneity, that $z=1$, and then we get,
using the fact that $\vert x\vert \ge 1$,
\begin{eqnarray*}
{x^2+{1\over x^2}+1\over x^4+{1\over x^4}+1} \delta
(x+\delta^{-1}{1\over x})^2 &\ge& \delta {x^2(x^2+{1\over
x^2}+1)\over {x^4+{1\over x^4}+1}}
\\
&\ge& \delta>1.
\end{eqnarray*}
Thus, we can find a neighborhood $\tilde V_-$ of
$\overline{C_-\cap\BP_\BR^2}$, in $\BP_\BR^2$, such that
\begin{eqnarray}\label{f.3.n.3}
\vert r({R(X)\over \|R(X)\|})\vert \ge \vert r({X\over
\|X\|})\vert, \;\;\; \forall X\in \BR^3, \;\hbox{s.t. } \pi(X)\in
\tilde V_-.
\end{eqnarray}
Using formula (\ref{f.3.n.0}) we see that
$$
f^{-1}(C)\subset C\cup D.
$$
But since $f(D)=[-\delta^{-1},-\delta,1]\in C_+$, and since
$f([x,y,z])=[x^2,y^2,z^2]$ for $[x,y,z]\in C$, we see that
$f^{-1}(C_-)\subset C_-$. Considering the map $f$ only on
$\BP^2_\BR$, we see that we can find a neighborhood $V_-$ of
$\overline{C_-\cap\BP^2_\BR}$, in $\BP_\BR^2$, such that
$f^{-1}(V_-)\subset \tilde V_-$. Since $V_-\cup V_+$ is a
neighborhood of $C\cap \BP_\BR^2$, we can find $\epsilon$ small
enough such that $V_\epsilon \subset V_-\cup V_+$. From this we
can deduce
\begin{eqnarray}
\label{f.3.n.4} f^{-1}(V_\epsilon \setminus V_+)\subset
V_\epsilon.
\end{eqnarray}
Indeed, if $X$ is in $V_\epsilon\setminus V_+$ then $X\in V_-$,
and thus $f^{-1}(\{X\})\subset \tilde V_-$. If $Y$ is in
$f^{-1}(\{X\})$, then by (\ref{f.3.n.3})
$$
\epsilon \ge \vert r({X\over \|X\|})\vert=\vert r({R(Y)\over
\|R(Y)\|})\vert \ge \vert r({Y\over \|Y\|})\vert.
$$
Thus, $Y$ is in $V_\epsilon$.
 \ali
This implies that if $\epsilon$ is small enough so that
$(x_0,y_0)\not\in V_\epsilon$, then the iterates $f^n((x_0,y_0))$
do not enter the set $V_\epsilon$. Indeed, otherwise we can
consider the first entrance time $n_0$ into $V_\epsilon$. The
point $f^{n_0}((x_0,y_0))$ is necessarily in $V_\epsilon\setminus
V_+$ since $(x_0,y_0)$ is not in the attractive basin of $x_+$.
This implies that $f^{n_0-1}((x_0,y_0))$ is in $V_\epsilon$, which
is impossible.$\Box$

Denote by $P^\infty_\BR$ the line at infinity
$P^\infty_\BR=\{[x,y,0], (x,y)\in\BR^2\}$. Remark that the
restriction of $f$ to $P_\BR^\infty$ is given by
$$
f([x,y,0])=[x,\delta y,0].
$$
and that $P^\infty_\BR$ is backward invariant, and
$f^{-1}([x,y,0])=[x,\delta^{-1}y,0]$. This implies that there
exists an integer $N_0$ such that for any $X$ in
$\overline{P^\infty_\BR\setminus V_\epsilon}$ the point
$f^{-N_0}(X)$ is in $V_\epsilon$. (This comes from the fact that
$V_\epsilon$ contains a neighborhood of $x_+$ and $x_-$). Thus, we
can find a neighborhood $V_\infty$ of
$\overline{P^\infty_\BR\setminus V_\epsilon}$ such that
$f^{-N_0}(V_\infty)\subset V_\epsilon$. We may as well take
$V_\infty$ such that $f^n((x_0,y_0))\not\in V_\infty$ when $n\le
N_0$. Let us now prove that the iterates $f^n((x_0,y_0))$ cannot
enter the set $V_\infty$: indeed, if $f^n((x_0,y_0))\in V_\infty$
then $n\ge N_0$ and thus $f^{n-N_0}$ is in $V_\epsilon$ which is
in contradiction with the result we proved above.
Hence, we proved that the iterates $f^n((x_0,y_0))$ do
not enter neither $V_\epsilon$ nor $V_\infty$. This concludes the
proof of the lemma since
$V_\epsilon \cup V_\infty$ contains a neighborhood of
$P^\infty_\BR$.$\Box$
 \ali

We consider the operator $\tilde \Gamma_{\nn, \lambda}$ in
$Sl_2(\BR)$ defined by
$$ \tilde \Gamma_{\nn, \lambda} = (\delta)^{\demi}
(D_{\delta})^{-1} (D_{\sqrt{\delta}})^{-n} \Gamma_{\nn,\lambda}
D_{\sqrt{\delta}}^n,
$$
and we set
$$
\Pi_\nn(\lambda)= \prod_{k=0}^{n-1} (\sqrt{\delta}
a(\gamma^k\lambda)+\sqrt{\delta}^{-1} d(\gamma^k\lambda)).
$$      
From (\ref{f.2.2}) and (\ref{f.2.4}) we have
\begin{eqnarray*}
\Gamma_{\nn, \lambda}=
\left(
\begin{array}{cc}
a_\nn(\lambda) & b_\nn(\lambda)
\\
c_\nn(\lambda) &
d_\nn(\lambda)
\end{array}
\right)&=&
\left(
\begin{array}{cc}
a_\nn(\lambda) &\alpha^{-n} b(\gamma^{n}\lambda)
\\
\alpha^n c(\gamma^n \lambda) &
d_\nn(\lambda)
\end{array}
\right)
\\
&=&
\left(
\begin{array}{cc}
a_\nn(\lambda)  & \sqrt{\delta}^{-n} \Pi_\nn(\lambda) b(\lambda)
\\
\sqrt{\delta}^{n} \Pi_\nn(\lambda) c(\lambda) &
d_\nn(\lambda)
\end{array}
\right) 
\end{eqnarray*}     
thus, we get
\begin{eqnarray} \label{f.3.1}
\tilde \Gamma_{\nn, \lambda}= 
\left(
\begin{array}{cc}
\sqrt{\delta} a_\nn(\lambda) & \sqrt{\delta} b(\lambda)
\Pi_\nn(\lambda)
\\
\sqrt{\delta}^{-1} c(\lambda) \Pi_\nn(\lambda) &
\sqrt{\delta}^{-1} d_\nn(\lambda)
\end{array}
\right).
\end{eqnarray}
Since $\tilde \Gamma_{\nn, \lambda}$ is in $Sl_2(\BR)$, we deduce
from lemma \ref{l.3.1} that
\begin{eqnarray} \label{f.3.3}
\vert \Pi_\nn(\lambda)\vert \le C_2
\end{eqnarray}
where $C_2={\sqrt{1+C_1^2}\over \sqrt{\vert
b(\lambda)c(\lambda)\vert }}$ is a constant depending only on
$\delta$, $\lambda$ (we know that $b(\lambda)$ and $c(\lambda)$
cannot be null, from proposition \ref{p.2.1}, since otherwise
$\phi(\lambda)$ would be in the attractive bassin of $x_+$). We
also deduce from lemma \ref{l.3.1}, formula (\ref{f.3.1}) and
relation (\ref{f.3.3}) that there exists a constant $C_3$ such
that for any $n$
\begin{eqnarray} \label{f.3.4}
C_3^{-1}\|X \|^2 \le \| \tilde \Gamma_{\nn, \lambda} X \|^2 \le
C_3 \|X\|^2,
\end{eqnarray}
for any vector $X$ in $\BR^2$ (and where $\|\;\|$ is the usual
norm in $\BR^2$).
\begin{lem} \label{l.3.2}
For any $\lambda$ in $\supp\mu$ there exists a constant $C_4>0$
and a subsequence $n_k$ such that for any solution $f$ of
$(E_\lambda)$ on $I_{<n_k+1>}$ we have
\begin{eqnarray} \label{f.3.5}
C_4^{-1} \le {\int_{I_{<n_k>}} \vert f\vert^2 dm_{n_k} \over
\int_{I_{<n_k+1>}\setminus I_{<n_k>} } \vert f\vert^2
dm_{<n_k+1>}}\le C_4.
\end{eqnarray}
\end{lem}
Proof: Remark that it is enough to prove this result for
$\w=(1,\ldots ,1, \ldots)$ since the measure $m_\nn(\w')$ is just,
up to a scalar factor, the image of $m_\nn(\w)$  by the right
composition of $\Psi_i$ and $\Psi_i^{-1}$ that send the interval
$I_\nn(\w)$ to $I_\nn(\w')$. Hence, we take $\w=(1, \ldots
,1,\ldots)$ in this lemma.

We first introduce the bilinear form $K_{\nn,\lambda}:\BR^2\times
\BR^2\rightarrow \BR$ defined as follows: for $(X,Y)$ in
$\BR^2\times \BR^2$ we denote by $f$ and $g$ the solutions of
$(E_\lambda)$ with initial conditions
$$
\left(\begin{array}{c} f(0) \\
f'(0)
\end{array}
\right) =X, \;\;\;
\left(\begin{array}{c} g(0) \\
g'(0)
\end{array}
\right)=Y
$$
and we set
$$
K_{\nn,\lambda}(X,Y)= \int_{I_\nn} fg dm_\nn.
$$
Clearly $K_{\nn,\lambda}$ is a positive definite symmetric
bilinear form. We simply write $K_{\nn,\lambda}(X)$ for $K_{\nn,
\lambda}(X,X)$ (it is possible to give an explicit expression of
$K_{\nn,\lambda}$ in terms of the propagator and its derivative
but we will not need it).

Take a solution $f$ of $(E_\lambda)$ on $I_{<n+1>}$. We denote by
$\tilde f$ the function on $I_\nn$ equal to the "pull-back" of
$f_{|I_\nun\setminus I_\nn}$ to $I_\nn$, i.e.
$$
\tilde f= f_{|I_\nun\setminus I_\nn} \circ \Psi_1^{-(n+1)}\circ
\Psi_2\circ \Psi_1^{n}.
$$
Clearly
$$
\left(\begin{array}{c} \tilde f(0) \\
\tilde f'(0)
\end{array}
\right) = (D_\delta)^{-1} \Gamma_{\nn,\lambda}
\left(\begin{array}{c} f(0) \\
f'(0)
\end{array}
\right),
$$
and
$$
\int_{I_\nun} \vert f\vert^2 dm_\nun =\int_{I_\nn} \vert f\vert^2
dm_\nn + \delta^{-1}\int_{I_\nn} \vert \tilde f\vert^2 dm_\nn.
$$
This can be translated in
$$
K_{\nun,\lambda}(X)=K_{\nn,\lambda}(X)+\delta^{-1}
K_{\nn,\lambda}(D_\delta^{-1} \Gamma_{\nn,\lambda} X).
$$
Define now
$$
\tilde K_{\nn,\lambda}(X)= K_{\nn,\lambda}(D_{\sqrt{\delta}}^n X).
$$
We see that the previous relation translates in
\begin{eqnarray} \label{f.3.7}
\tilde K_{\nun,\lambda} ( X)= \tilde
K_{\nn,\lambda}(D_{\sqrt{\delta}}X)+ \tilde K_{\nn,\lambda}(\tilde
\Gamma_{\nn,\lambda}(D_{\sqrt{\delta}} X)).
\end{eqnarray}
Take $\lambda$ in $\supp\mu$, we first prove that there exists a
constant $C_5$ and a sequence $n_k$ such that
\begin{eqnarray}\label{f.3.8}
{\sup_{\|X\|=1} \tilde K_{<n_k>,\lambda}(X) \over \inf_{\|X\|=1}
\tilde K_{<n_k>,\lambda}(X)}\le C_5.
\end{eqnarray}
This comes from the following technical lemma
\begin{lem}\label{l.3.3}
If $K$ is a positive quadratic form on $\BR^2$ and $\Gamma$ an
element of $Sl_2(\BR)$ with $\vert \tr \Gamma \vert < 2$ then we
have the inequality
$$
(\sup_{\|X\|=1} K(X))\left( {1-\vert \tr \Gamma\vert^2/4\over
\|\Gamma \|^2} \right)^2
 \le K(Z)+K(\Gamma Z) \le (\sup_{\|X\|=1} K(X) )(1+\|\Gamma\|^2).
 $$
for all $Z$ in $\BR^2$ with $\|Z\|=1$. (In the last formula
$\|\Gamma\|^2=\tr \Gamma^*\Gamma$).
\end{lem}
Proof: The inequality of the right hand side is trivial. For the
left hand side inequality  we may as well suppose that $K$ is
diagonal, i.e. that $K$ is of the form $K(X)=\rho_1 x^2 +\rho_2
y^2$ for
 $X=\left(\begin{array}{c} x\\y\end{array}\right)$  (indeed, by an
orthogonal change of variables $O$, the operator $O^*\Gamma O$
remain in $\Sl_2(\BR)$ and keeps the same trace). We suppose that
$\rho_1\ge \rho_2$. If
$$
\Gamma=\left(\begin{array}{cc} a&b\\c&d
\end{array}\right)
$$
and $Z=\left(\begin{array}{c} x\\y\end{array}\right)$, with
$x^2+y^2=1$, then
\begin{eqnarray*}
K(Z)+K(\Gamma Z)&\ge& \rho_1( x^2 +(ax+by)^2)
\\
&\ge & \rho_1 {1\over \| \left(\begin{array}{cc} 1 & 0
\\a&b\end{array}\right)^{-1}\|^2}
\\
&\ge& \rho_1 {b^2\over \|\Gamma\|^2}.
\end{eqnarray*}
Suppose now that $ad\le 0$ then $\vert bc\vert \ge 1$ and thus
$\vert b\vert^2 \ge {1\over \|\Gamma\|^2}$.
 \ali
Suppose now that $ad\ge 0$ then $ ad\le \tr(\Gamma)^2/4$ thus
$\vert bc\vert \ge (1-\tr(\Gamma)^2/4)$ and we get the left hand
side inequality.$\Box$
 \ali\ali
We apply this result to $K_{\nn,\lambda}$. Remark first that $\tr
(\Gamma_{\nn,\lambda})=\sqrt{\delta} a_\nn(\lambda)
+\sqrt{\delta}^{-1}d_\nn(\lambda)$. From the definition of
$\Pi_\nn(\lambda)$ and relation (\ref{f.3.3}), we deduce that
there exists a subsequence $n_k$ such that $\vert \tr \tilde
\Gamma_{<n_k-1>,\lambda}\vert \le {2\over \sqrt{3}}$, for example.
From relation (\ref{f.3.7}) and the previous lemma we easily
deduce relation (\ref{f.3.8}) with $C_5={3\over 2}\delta
C_3^2(1+C_3)$.

We are now in position to conclude the proof of lemma \ref{l.3.2}.
Consider a function $f$, solution of $(E_\lambda)$ on
$I_{<n_k+1>}$. Set
$$
X=D_{\sqrt{\delta}}^{-n}\left(\begin{array}{c} f(0) \\ f'(0)
\end{array}\right).
$$
By definition we have
\begin{eqnarray*}
&& \int_{I_{<n_k>}} \vert f\vert^2 dm_{<n_k>}=\tilde
K_{<n_k>,\lambda}(X),\;\; \hbox{ and } \\
&& \int_{I_{<n_k+1>}\setminus I_{<n_k>} } \vert f\vert^2
dm_{<n_k+1>}=\tilde K_{<n_k>,\lambda}(\tilde
\Gamma_{<n_k>,\lambda} X).
\end{eqnarray*}
From relation (\ref{f.3.8}) and relation (\ref{f.3.4}) we see that
the inequality of lemma \ref{l.3.2} is satisfied for $C_4=C_3^2
C_5$.$\Box$

We can now finish the proof of lemma \ref{l.noteigenvalue}. Take any
blow-up $\w$. Suppose that $\lambda$ in $\supp\mu$ is an
eigenvalue of $H_\infi^\pm(\w)$ with eigenfunction $f$, $\|
f\|=1$. For any $\epsilon >0$ there exists $N$ such that
$$
\int_{I_\infi\setminus I_{<N>}} \vert f\vert^2 dm_\infi \le
\epsilon.
$$
Clearly, this contradicts lemma \ref{l.3.2} for $\epsilon$ small
enough. This concludes the proof of theorem
\ref{t.3.1}.$\diamondsuit$
\vfill\break
\centerline{\begin{picture}(0,0)%
\includegraphics{carte.pstex}%
\end{picture}%
\setlength{\unitlength}{2368sp}%
\begingroup\makeatletter\ifx\SetFigFont\undefined%
\gdef\SetFigFont#1#2#3#4#5{%
  \reset@font\fontsize{#1}{#2pt}%
  \fontfamily{#3}\fontseries{#4}\fontshape{#5}%
  \selectfont}%
\fi\endgroup%
\begin{picture}(12762,14205)(76,-13410)
\put(8251,-8760){\makebox(0,0)[lb]{\smash{\SetFigFont{7}{8.4}{\familydefault}$D=\{q_1+\delta^{-1}q_2=0\}$
}}}
\put(6751,-10110){\makebox(0,0)[lb]{\smash{\SetFigFont{7}{8.4}{\familydefault}$D'=\{q_1+\delta^{-2}q_2=2\delta^{-1}q\}$
}}}
\put(12676,-5535){\makebox(0,0)[lb]{\smash{\SetFigFont{7}{8.4}{\familydefault}$x_-=[1,0,0]$
}}}
\put(4876,-7935){\makebox(0,0)[lb]{\smash{\SetFigFont{7}{8.4}{\familydefault}$[-\delta^{-1},-\delta,-1]$
}}}
\put(6676,615){\makebox(0,0)[lb]{\smash{\SetFigFont{7}{8.4}{\familydefault}$x_+=[0,1,0]$
}}}
\put(10951,-1260){\makebox(0,0)[lb]{\smash{\SetFigFont{7}{8.4}{\familydefault}$\{q=0\}$
}}}
\put(9751,-4936){\makebox(0,0)[lb]{\smash{\SetFigFont{7}{8.4}{\familydefault}${\cal C}_-$
}}}
\put(3001, 89){\makebox(0,0)[lb]{\smash{\SetFigFont{7}{8.4}{\familydefault}$l=[1,-\delta,0]$
}}}
\put(5401,-13410){\makebox(0,0)[lb]{\smash{\SetFigFont{7}{8.4}{\familydefault}Picture of $\hbox{I\!\!\! P}^2_{\Bbb R}$ for $\delta=2$
}}}
\put( 76,-5536){\makebox(0,0)[lb]{\smash{\SetFigFont{7}{8.4}{\familydefault}$x_-$
}}}
\put(6376,-11761){\makebox(0,0)[lb]{\smash{\SetFigFont{7}{8.4}{\familydefault}$x_+$
}}}
\put(9151,-11161){\makebox(0,0)[lb]{\smash{\SetFigFont{7}{8.4}{\familydefault}$l$
}}}
\put(1201,-6136){\makebox(0,0)[lb]{\smash{\SetFigFont{7}{8.4}{\familydefault}$V_-$
}}}
\put(11026,-5611){\makebox(0,0)[lb]{\smash{\SetFigFont{7}{8.4}{\familydefault}$V_-$
}}}
\put(6076,-10786){\makebox(0,0)[lb]{\smash{\SetFigFont{7}{8.4}{\familydefault}$V_+$
}}}
\put(6151,-61){\makebox(0,0)[lb]{\smash{\SetFigFont{7}{8.4}{\familydefault}$V_+$
}}}
\put(1576,-2236){\makebox(0,0)[lb]{\smash{\SetFigFont{7}{8.4}{\familydefault}$V_\infty$
}}}
\put(11176,-8536){\makebox(0,0)[lb]{\smash{\SetFigFont{7}{8.4}{\familydefault}$V_\infty$
}}}
\put(6901,-960){\makebox(0,0)[lb]{\smash{\SetFigFont{7}{8.4}{\familydefault}${\cal C}$
}}}
\put(6976,-1635){\makebox(0,0)[lb]{\smash{\SetFigFont{7}{8.4}{\familydefault}${\cal C}_+$
}}}
\put(7876,-4260){\makebox(0,0)[lb]{\smash{\SetFigFont{7}{8.4}{\familydefault}$[1,1,-1]$
}}}
\end{picture}
}


\begin{thebibliography}{100}
\bibitem{CarmonaL}
R. CARMONA and J. LACROIX, Spectral Theory of Random Schr\"odinger
Operators, Probabilities and Applications, Birkha\"user, Boston,
1990.
\bibitem{FSibony}
J. E. FORNAESS and N. SIBONY, {\it Complex dynamics in higher
dimension II.} In Modern Methods in Complex Analysis (Princeton, NJ,
1992), 135--182, Ann. of Math. Stud., 137, Princeton Univ. Press,
Princeton, NJ, 1995.
\bibitem{Fukushima1}
M. FUKUSHIMA, Y. OSHIMA and M. TAKEDA, Dirichlet forms and
Symmetric Markov Processes, de Gruyter Stud. Math. 19, Walter de
Gruyter, Berlin, New-york, 1994.
\bibitem{Fukushima2}
M. FUKUSHIMA, {\it Dirichlet forms, diffusion processes and
spectral
 dimensions
for nested fractals,} in : Ideas and Methods in Mathematical
Analysis, Stochastics and Applications, Proc. Conf. in Memory of
Hoegh-Krohn, vol. 1 (S. Albevario et al., eds.), Cambridge Univ.
Press, Cambridge, 1993, pp. 151-161.
\bibitem{GriffithsH}
P. GRIFFITS, J. HARRIS, Principles of algebraic geometry.
Wiley Classics Library. John Wiley \& Sons, Inc., New York, 1994.
xiv+813 pp
\bibitem{Hormander}
L. H\"ORMANDER, Notions of Convexity. Progress in Mathematics,
127. Birkhäuser Boston, Inc., Boston, MA, 1994. viii+414 pp.
\bibitem{Kigami1}
J. KIGAMI,  {\it Harmonic calculus on p.c.f. self-similar sets,}
Trans. Am. Math. Soc., 335:721-755, 1993.
\bibitem{KigamiL}
J. KIGAMI and M. L. LAPIDUS, {\it Weyl's problem for the spectral
distribution of Laplacians on p.c.f. self-similar fractals,}
Commun. Math. Phys. 158, no. 1, (1993), 93-125.
\bibitem{PasturF}
L. PASTUR and A. FIGOTIN,
Spectra of Random and Almost-Periodic Operators,
Grundlehren der mathematischen Wissenschaften,
297, Springer-Verlag, Berlin Heidelberg 1992.
\bibitem{Sabot1}
C. SABOT, {\it Existence and uniqueness of diffusions on finitely
ramified self-similar fractals,} in Ann. Scient. Ec. Norm. Sup.,
4\`eme s\'erie, t. 30, 1997, pp. 605 \`a 673.
\bibitem{Sabot3}
C. SABOT,
{\it
Integrated density of states of self-similar Sturm-Liouville
operators
and holomorphic dynamics in higher dimension},
Ann. Inst. H. Poincar\'e  Probab.
Statist 37 (2001), no. 3, 275-311.
\bibitem{Sabot5}
C. SABOT, Spectral properties of self-similar lattices and
iteration of rational maps, M\'emoires de la SMF, 92 (2003),
arXiv.org/math-ph/0201040.
\bibitem{Sabot6} C. SABOT, {\it Laplace operators on
fractal lattices with random blow-up}, in Potential Analysis, 20,
177-193, 2004.
arXiv.org/math-ph/0201041.
\bibitem{Sabot-review1}
C. SABOT,
{Electrical networks, symplectic reductions, and applications
to the renormalization map of self-similar lattices.},
to appear in Proc. of Symp. in Pure Math., Mandelbrot Jubilee.
arXiv/math-ph/0304015.
\bibitem{Sibony1}
N. SIBONY, {\it Dynamique des applications rationnelles de
$\bold P\sp k$} (French). Dynamique et G\'eom\'etrie Complexes (Lyon,
1997), ix--x, xi--xii, 97--185, Panor. Synth\`eses, 8, Soc. Math.
France, Paris, 1999.
\end{thebibliography}
\end{document}